      [1999/12/01 v1.4c Il Nuovo Cimento]
\documentclass{cimento}

\usepackage{graphicx}  
\title{Testing different stellar mass estimators at $1<z<2$}
\author{M.~Longhetti\from{ins:bre}\ETC,
P.~Saracco\from{ins:bre}
\atque
A.~Mignano\from{ins:bre}
}
\instlist{\inst{ins:bre} INAF - Osservatorio Astronomico di Brera Milano - Italy
                          }
\PACSes{
\PACSit{98.52.Eh}{Elliptical galaxies}
\PACSit{95.75.Fg}{Spectroscopy and spectrophotometry}
}
 
\begin{document}

\maketitle

\begin{abstract}
Physical parameters of galaxies (as luminosity, stellar mass, age) are often 
derived by means of the model templates which best fit their spectro-photometric data. 
We have performed a quantitative test aimed at exploring the ability of this 
procedure in recovering the physical parameters of early-type galaxies at $1<z<2$. 
A wide range of simulated SEDs, reproducing those of early-type galaxies 
at $1<z<2$ with assigned age and mass, are used to build mock photometric catalogs 
with wavelength coverage and photometric uncertainties similar to those of two 
topical surveys (i.e. VVDS and GOODS). The best fitting analysis of the simulated 
photometric data allows to study the differences among the recovered parameters 
and the input ones. Results indicate that the stellar masses measured by means of 
optical bands are affected by larger uncertainties with respect to those obtained 
from near-IR bands, and they frequently underestimate the real values. The M/L 
ratio in the V band results strongly underestimated, even when derived from the 
recently proposed recipe based on rest-frame optical colours (e.g. (B-V)).
\end{abstract}

\section{Mock photometric catalogs}
As a first step aimed at exploring the ability of different mass estimators in
recovering the stellar mass content of early-type galaxies at $1<z<2$,
we built a set of mock photometric catalogs.  We adopted the BC03 \cite{ref:1} code,
assuming Salpeter IMF and solar metallicity, to generate the SEDs of a wide family
of models reproducing the observations of early-type galaxies. The selected SF histories are
described by an exponentially declining SFR with time scale $\tau=0.6$ Gyr,
and by the superimposition of $\tau=0.1$ Gyr models at different times 
simulating secondary bursts involving between 5\% and 20\% of the total final mass.
The relevant range of redshift
is described distributing the model templates at $z=1.0,1.5,2.0$. The chosen combinations
among ages, redshift and times of the secondary burst constitute 57 template spectra, 
among which 11 with no secondary
star forming episodes and 16 with recent (i.e.  $<1$ Gyr) starburst,
half of which forms 5\% of the total final mass while the contribution of the other half
is 20\%.
Finally, each generated SED is proposed with three possible values of dust extinction
A$_{V}=0.0,0.2,0.5$, assuming the reddening law of Calzetti et al. \cite{ref:2}.

\section{Analysis}
Among a wide set of templates, we searched for the one best fitting each simulated
galaxy by means of the photometric redshift code {\em hyperz}
\cite{ref:3}.
The set of SEDs adopted to find the best fit template for each galaxy is composed by
models with SFR time scales $\tau=0.1,0.3,0.6,1.0$ Gyr
at solar metallicity, generated by means of the BC03 \cite{ref:1} code 
assuming Salpeter IMF.
In the best fitting procedure the extinction
has been allowed to vary between A$_{V}=0.0$ and A$_{V}=0.5$ 
and at each $z$ ages have been
forced to be lower than the Hubble time at that $z$.
Assuming that the redshift is known
within 0.1, we run the code with $z$ varying between +/- 0.05 of its real value.
Further details on the construction and analysis of the mock catalogs
can be found in a forthcoming paper (Longhetti et al. 2008 \cite{ref:4}).

The assumed set of parameters reproduces the choices generally made to study the
photometric properties of real early-type galaxies  with known spectroscopic redshift
(e.g.  \cite{ref:5}), since they include all the possible combinations of ages and
dust compatible with this class of galaxies, excluding strong dust reddened star
forming galaxies.

Once a best fit template has been associated to each simulated galaxy, we derived the following
physical parameters:

\noindent
$\bullet$ {\bf M$_{K}$} and {\bf M$_{K_{4.5}}$} = absolute K band magnitude derived from the observed K 
magnitude and from the observed IR flux at 4.5$\mu$m respectively, obtained applying 
the k-corrections calculated on the best fit template and the dust correction derived for the 
best fit value of A$_{V}$;

\noindent
$\bullet$ {\bf M$_{Kraw}$} = absolute K band magnitude derived from the observed K magnitude 
simply applying a constant k-correction depending only from the redshift, 
without any dust correction;

\noindent
$\bullet$ {\bf M$_{V_{J}}$} = absolute V band magnitude derived from the observed J magnitude, 
obtained applying the k-correction calculated on the best fit template and the dust 
correction derived for the best fit value of A$_{V}$;

\noindent
$\bullet$ {\bf $\mathcal{M/L_{K}}$} and {\bf $\mathcal{M/L_{V}}$} = mass to light ratio in the K and V bands 
respectively, as derived from the best fit template;

\noindent
$\bullet$ {\bf $\mathcal{M/L_{V}}$(B-V)} = mass to light ratio in the V band derived as
$\log[(\mathcal{M/L})_{V}]=-0.628+1.305\ (B-V)_{0}$ by
Bell et al. \cite{ref:6}, where the (B-V) rest-frame colour is derived from the best fit template;

\noindent
$\bullet$ {\bf $\mathcal{M}$(b)} = stellar mass content derived from the normalization factor needed 
to scale the model templates to match on average the observed available fluxes 
(b parameter in the hyperz code);

\noindent
$\bullet$ {\bf $\mathcal{M}_{K}$}, {\bf $\mathcal{M}_{V_{J}}$} and {\bf $\mathcal{M}_{V_{J}}$(B-V)} = 
stellar mass content derived 
from the absolute M$_{K}$ and M$_{V_{J}}$ magnitudes and assuming the mass to light ratios 
$\mathcal{M/L_{K}}$, $\mathcal{M/L_{V}}$ and $\mathcal{M/L_{V}}$(B-V) respectively;

\noindent
$\bullet$ {\bf M$_{V_{J}}^{A_{V}=0}$}, {\bf $\mathcal{M}_{V_{J}}$(B-V)$^{A_{V}=0}$} and 
{\bf $\mathcal{M/L_{V}}$(B-V)$^{A_{V}=0}$} 
= parameters obtained on the subsample of the simulated galaxies with no 
dust extinction imposing A$_{V}=0$ in hyperz (i.e., not affected by dust uncertainties).

Then the derived physical parameters of the simulated galaxies have been
compared with their real known values (and a more detailed analysis can be
found in \cite{ref:4}).

\section{Results}
\begin{figure}
\centering
\includegraphics[width=5.7truecm]{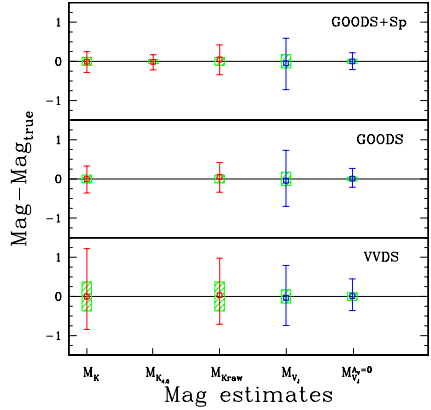}
\caption{Uncertainties in the estimates of the absolute magnitudes in the K and V band.
The reported errorbar represents the whole range of values spanned by the difference
between the real and the recovered magnitudes. The square dot is the average value
of the same difference. The shaded areas highlight the range of uncertainties of the
apparent magnitudes used as the starting point to derive the absolute ones.
}
\end{figure}
As far as the {\bf luminosities} are concerned, from Fig. 1 it can be seen that
M$_{K}$ generally recovers within 0.2-0.3 magnitudes the real value
in the case of the GOODS mock catalog \cite{ref:7}
($\Delta$K$<0.1$ for all the simulated galaxies, K$_{lim}\simeq21.5$)
while due to larger errors in the apparent magnitudes
($\Delta$K=0.2) the uncertainty is around 0.5 magnitudes in the case of the 
VVDS mock catalog \cite{ref:8} (K$_{lim}\simeq20.5$), and it
can raise to more than 1.0 magnitude in the case of the fainter
objects (i.e.  with $\mathcal{M}_{star}=0.3\times10^{11}\mathcal{M}_{\odot}$ at $z=2$ for
which K$\approx$21.5). 
Spitzer IR data with small
errors ($\Delta$(4.5$\mu$m)=0.05) allow to obtain estimates of the K magnitudes even better than 0.2 magnitudes.
It is particularly remarkable that the K$_{raw}$ values are well determined within
0.3-0.4 magnitudes when the apparent K magnitudes are affected by small errors ($\Delta$K$<0.1$).
This demonstrates the easiness to derive a reliable estimate of the near-IR luminosities
for early-type galaxies at known redshift.
In the same figure, we have reported the estimate of the V band absolute magnitude M$_{V_{J}}$ that
displays a much larger uncertainty than that of the K band one, partly because
of large errors in the J band
also for the GOODS mock catalog ($\Delta$(J)=0.15-0.20), and partly because of 
the strong dust effects expected at $\lambda\simeq5500 \AA$.
\begin{figure}
\centering
\includegraphics[width=8.9truecm]{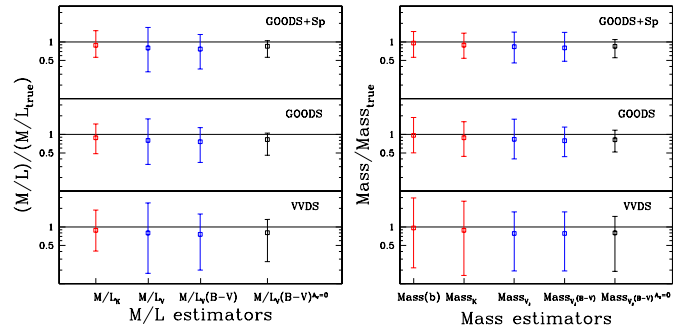}
\caption{Retrieved mass to light ratios M/L in the K and V bands (left panel)
and mass values (right panel)
are reported as normalized to their true values  (see previous
section for detailed definitions}.
\end{figure}
In Fig. 2 (left panel), the {\bf $\mathcal{M/L}$ ratios} in the K and V bands are reported
normalized to their true values.
The retrieved values of $\mathcal{M/L}_{K}$ are well within a factor $\pm2$ from the real values, and only
for the fainter galaxies in the VVDS mock catalog
the uncertainties grow up to 0.4-1.9. For both the two catalogs (even when analyzed
with the support of IR Spitzer data) it can be noted a trend to
underestimate the real value of $\mathcal{M/L}_{K}$ by a factor of about 0.9.
The underestimate of the $\mathcal{M/L}$
value in the V band is even larger (i.e. around 0.8), and the uncertainties are
much larger in this band than in the K one (i.e. ($\mathcal{M/L}_{V}$)/($\mathcal{M/L}_{V}$)$^{true}$
within 0.3-1.7 for the smaller errors).
The values of $\mathcal{M/L}_{V}$(B-V) 
show a smaller range of uncertainties with respect to $\mathcal{M/L}_{V}$,
but the underestimate is larger.
Part of this effect is due to the dust extinction, but even when
only no-dust models are considered $\mathcal{M/L}_{V}$(B-V)$^{A_{V}=0}$ is 
around 0.8 times the real value
 and the resulting uncertainty range is 0.5-1.0
for the GOODS mock catalog.
Since Bell et al. \cite{ref:6} do not explicitly state
at which mass they refer in their definition of the $\mathcal{M/L}$ ratio,
it is possible that their proposed formula has to be considered for
the mass still locked into stars (as often assumed when dealing with the stellar content of
galaxies) that is expected to be around
0.7-0.8 of that adopted in the present paper (resulting from the integration
of the SFR over the time up to the age of the model).

\noindent
On the right panel of Fig. 2, the retrieved {\bf mass} values obtained by means of different estimators
are reported normalized to the true values.
The mass estimate obtained by means of the template scaling factor (e.g., $b$ parameter of {\em hyperz})
is the better determined with respect of any other estimator (\cite{ref:4}).
The values of $\mathcal{M}_{K}$ 
are those which closer approach the previous ones, even if on average
they slightly underestimate the real values.
The trend towards the underestimate of the masses is much more marked when
calculated with the V luminosities
and both $\mathcal{M/L}_{V_{J}}$ and $\mathcal{M/L}_{V_{J}}$(B-V), for which
the recovered masses are only 70-80\% of the real values on average,
and the dispersions around the real values are much larger than the previous cases.

Summarizing, when spectro-photometric data are used to find the best fit templates reproducing the observed
early-type galaxies, 
it is easier to obtain a good determination
of near-IR absolute luminosities than optical ones, and
the best mass estimate, not affected by any systematic trend, is
that derived by the scaling factor between templates and data on average. Alternatively,
the near-IR bands can be safely used to derive the mass content of early-type galaxies, while
optical bands produces much larger uncertainties and general underestimate of the stellar
masses which are difficult to be taken into account. 

\acknowledgments
This research has received financial support from the
Istituto Nazionale di Astrofisica (Prin-INAF CRA2006 1.06.08.04).

\end{document}